# Giant magnetoresistance of Dirac plasma in high-mobility graphene

Na Xin[1,2+], James Lourembam[1+], P. Kumaravadivel[1,2+], A. E. Kazantsev[1], Zefei Wu[2], Ciaran Mullan[1], Julien Barrier[1,2], Alexandra A. Geim[2], I. V. Grigorieva[1], A. Mishchenko[1], A. Principi[1], V. I. Falko[1,2], L. A. Ponomarenko[3*], A. K. Geim[1,2,4*], Alexey I. Berdyugin[1,2,4,5*]

[1]Department of Physics & Astronomy, University of Manchester, Manchester M13 9PL, United Kingdom
[2]National Graphene Institute, University of Manchester, Manchester M13 9PL, United Kingdom
[3]Department of Physics, University of Lancaster, Lancaster LA1 4YW, United Kingdom
[4]Department of Materials Science and Engineering, National University of Singapore, 9 Engineering Drive 1, Singapore 117575, Singapore
[5]Department of Physics, National University of Singapore, 2 Science Drive 3, Singapore 117551, Singapore
[+] Those authors contributed equally to this work
[*]Correspondence related to this work should be addressed to A.I.B. (alexey@nus.edu.sg), L.A.P (l.ponomarenko@lancaster.ac.uk) and A.K.G. (geim@manchester.ac.uk)

**The most recognizable feature of graphene's electronic spectrum is its Dirac point around which interesting phenomena tend to cluster. At low temperatures, the intrinsic behavior in this regime is often obscured by charge inhomogeneity[1,2] but thermal excitations can overcome the disorder at elevated temperatures and create an electron-hole (e-h) plasma of Dirac fermions. The Dirac plasma has been found to exhibit unusual properties including quantum critical scattering[3–5] and hydrodynamic flow[6–8]. However, little is known about the plasma's behavior in magnetic fields. Here we report magnetotransport in this quantum-critical regime. In low fields, the plasma exhibits giant parabolic magnetoresistivity reaching >100% in 0.1 T even at room temperature. This is orders of magnitude higher than magnetoresistivity found in any other system at such temperatures. We show that this behavior is unique to monolayer graphene, being underpinned by its massless spectrum and ultrahigh mobility, despite frequent (Planckian-limit) scattering[3–5,9–14]. With the onset of Landau quantization in a few T, where the e-h plasma resides entirely on the zeroth Landau level, giant linear magnetoresistivity emerges. It is nearly independent of temperature and can be suppressed by proximity screening[15], indicating a many-body origin. Clear parallels with magnetotransport in strange metals[12–14] and so-called quantum linear magnetoresistance predicted for Weyl metals[16] offer an interesting playground to further explore relevant physics using this well-defined quantum-critical 2D system.**

A variety of mechanisms – both intrinsic and extrinsic – can lead to large magnetoresistance (MR) in metallic systems. The quest to understand those mechanisms has continued for longer than a century but many gaps still remain, which is especially obvious for MR reported in newcomer materials such as various Dirac and Weyl systems[17–25], strange metals[12–14], etc. The history and current status of the research field are briefly reviewed in Sections 1&2 of Methods. Whichever mechanism is behind a particular MR behavior, it always relies on bending of electron trajectories by magnetic field $B$ and, accordingly, high carrier mobility $\mu$ is an essential prerequisite for the observation of large MR. Colossal MR (reaching ~$10^6$ % in 10 T) was observed in a number of high-$\mu$ systems at liquid-helium temperatures[17–25]. However, because mobility decreases with increasing temperature $T$, this usually results only in a tiny MR at $T$ above liquid-nitrogen. Those few materials in which carriers remain highly mobile at room $T$ (such as doped graphene and InSb)[26–28] are all non-compensated systems and, in agreement with the classical theory of normal metals[29], their longitudinal resistivity $\rho_{xx}$ saturates in high $B$, leading again to little MR. Only the presence of extended defects[30–32] or a special design of 4-probe devices[26,33] which creates strongly nonuniform current flows can lead to large – but extrinsic – magnetoresistance (Methods).

As shown below, thermally excited charge carriers in monolayer graphene (MLG) at the neutrality point (NP) exhibit an anomalously high mobility $\mu$ exceeding 100,000 cm$^2$ V$^{-1}$ s$^{-1}$ at room $T$, despite the fact that the system is strongly interacting[3–8] and the electron-hole scattering time $\tau_P$ is ultimately short, being limited by the uncertainty principle $\tau_P^{-1} \approx C k_B T/h$ where $k_B$ and $h$ are the Boltzmann and Planck constants, respectively, and $C \approx 1$ is the interaction constant[3–5,9–12]. Importantly, unlike any known system with high $\mu$ at room $T$, the Dirac plasma is compensated (charge neutral) so that its zero Hall response allows non-saturating MR[29] whereas the high $\mu$ makes it colossal. To emphasize how unique magnetoresistivity $\rho_{xx}(B)$ of the Dirac plasma is, we provide its comparison with graphite (multilayer graphene[34]) and charge-neutral bilayer graphene (BLG), another quantum-critical system exhibiting Planckian scattering but having massive charge carriers with modest mobilities[9,10].

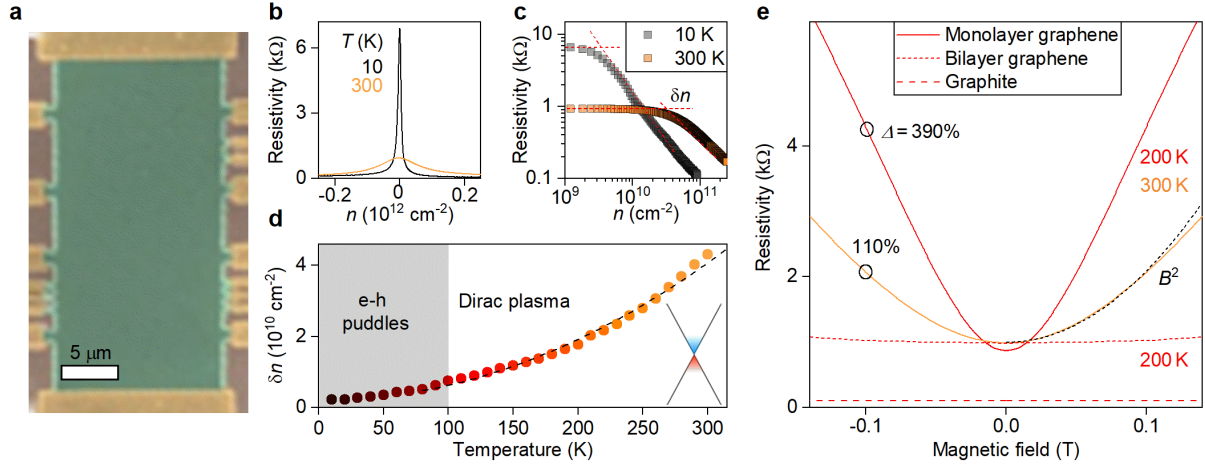

**Fig. 1| Electron transport in graphene's Dirac plasma. a**, Scanning electron micrograph of one of the studied MLG devices in false color. Green areas indicate encapsulated graphene intentionally misaligned with both top and bottom hBN to avoid superlattice effects, golden - metallic contacts, and brown - oxidized Si wafer serving as a gate. **b**, Zero-$B$ resistivity of MLG near the NP as a function of gate-induced carrier density. **c**, Data from panel **b** replotted in a double logarithmic scale to evaluate $\delta n$ as indicated by the dashed lines[9]. **d**, $\delta n$ as a function of $T$. Black curve: parabolic dependence. Above 100 K, $n_{th}$ in this device becomes several times higher than the residual charge inhomogeneity. Inset: schematics of the graphene spectrum with thermally excited carriers indicated in blue and red. **e**, Resistivity of the compensated Dirac plasma in small $B$ at representative $T$ (solid curves). Black curve: parabolic fit at 300 K. Black circles and values: $\Delta$ at 0.1 T. Short- and long- dash curves: resistivity of charge-neutral bilayer graphene and graphite, respectively, at the NP at 200 K. All the MLG data are from device D1. More examples of MR behavior for MLG, BLG and graphite are provided in Methods.

**Giant magnetoresistance in non-quantizing fields**

Our primary devices were multiterminal Hall bars made from MLG encapsulated in hexagonal boron nitride (Fig. 1a). We have studied several such devices and focus here on two of them (D1 and D2) showing representative behavior. At low $T$, their mobilities exceed $10^6$ cm$^2$ V$^{-1}$ s$^{-1}$ at characteristic carrier densities of $\sim 10^{11}$ cm$^{-2}$, being limited by edge scattering despite the devices' size $>10$ μm. Typical behavior of $\rho_{xx}$ as a function of the gate-induced density $n$ is shown in Fig. 1b. If the same curves are replotted on the log scale (Fig. 1c), it becomes clear that $\rho_{xx}$ responds to gate voltage only above a certain $n$ dependent on $T$. This behavior is commonly quantified as shown in Figs. 1c&d where $\delta n$ marks the gate-induced density that leads to notable changes in $\rho_{xx}$. At high $T$, the peak in $\rho_{xx}(n)$ broadens because of thermally excited electrons and holes in concentrations $n_{th} = (2\pi^3/3)(k_B T/hv_F)^2$ where $v_F$ is the Fermi velocity (Methods). The extracted $\delta n$ evolves $\propto T^2$ as expected (Fig. 1d) and its absolute value is $\sim 0.5 n_{th}$ which means that, to make changes in $\rho_{xx}$ visible on such log plots, gate-induced carriers are required in concentrations of $\sim 50\%$ of the thermal concentration. At low $T$, $\delta n$ saturates typically at $\sim 10^{10}$ cm$^{-2}$ because of residual charge inhomogeneity (e-h puddles of submicron scale)[1,2]. Below we focus on $T > 100$ K where thermal excitations totally overwhelm the residual $\delta n$.

The Dirac plasma's response to small fields is shown in Fig. 1e. One can see that $\rho_{NP} \equiv \rho_{xx}(n=0)$ increases proportionally to $B^2$, as expected from the classical Drude model[29]. However, the changes in $\rho_{NP}$ are unexpectedly large for this $T$ range. Indeed, if we consider 0.1 T as a characteristic field relevant for magnetic-sensor applications, then the relative magnetoresistivity $\Delta = [\rho_{xx}(B) - \rho_{xx}(0)]/\rho_{xx}(0)$ reaches ~110% at 300 K near the NP (Fig. 1e) and increases by a factor of 3-4 at 200 K. For comparison, $\Delta$ in normal metals rarely exceeds a small fraction of 1% above liquid-nitrogen temperatures. Even high-quality encapsulated bilayer, few-layer and multilayer graphene exhibit $\Delta(0.1T)$ reaching only ~1% at room $T$ (Methods). Also, the renowned giant MR based on spin flipping in ferromagnetic multilayers yields one-two orders of magnitude smaller changes in resistance[35,36] than those observed here for the Dirac plasma.

Further characterization of the e-h plasma is provided in Fig. 2. It shows that $\Delta$ rapidly diminishes away from the NP at characteristic densities $n \approx n_{th}$ (Fig. 2a). This is expected[29] because, for non-compensated systems, changes in $\rho_{xx}(B)$ should be small and saturate, if Hall resistivity $\rho_{xy} > \rho_{xx}$ (Methods). In contrast, for a charge-neutral systems (zero $\rho_{xy}$), the Drude model predicts non-saturating magnetoresistivity such that $\Delta = \mu_B^2 B^2$ where $\mu_B$ is the mobility in non-quantizing magnetic fields (Methods). The latter expression describes well the behavior observed in small $B$ (Fig. 2b). Fig. 2c plots the extracted $\mu_B$ as a function of $T$. The mobility exceeds 100,000 cm$^2$ V$^{-1}$ s$^{-1}$ at room $T$ and grows above 300,000 cm$^2$ V$^{-1}$ s$^{-1}$ below 150 K. Although high $\mu$ values are well known for the Fermi-liquid regime in doped graphene, it is unexpected that the mobility remains high in the presence of Planckian scattering, characteristic of the quantum-critical regime in neutral graphene[5,6]. For comparison, bilayer and multilayer graphene also exhibit very high mobilities at liquid-helium $T$ but their $\rho_{NP}(B)$ are practically flat at elevated $T$ (Fig. 1e), yielding $\mu_B$ of only ~10,000 cm$^2$ V$^{-1}$ s$^{-1}$ at 300 K (Extended Data Figs. 2&3). The dramatic difference in electronic quality between the e-h plasmas of relativistic and nonrelativistic fermions (in MLG and BLG, respectively) stems from the small effective mass $m$ characteristic of the Dirac spectrum ($\mu \propto m^{-1}$) and its low density of states which reduces the efficiency of electron scattering (Methods). Note however that the Dirac spectrum on its own is insufficient for achieving giant values of $\Delta$, and the high quality of MLG devices is paramount. This is emphasized by Extended Data Fig. 8 that shows magnetotransport for non-encapsulated graphene on a silicon oxide substrate. Such low-quality MLG exhibits three orders of magnitude smaller MR.

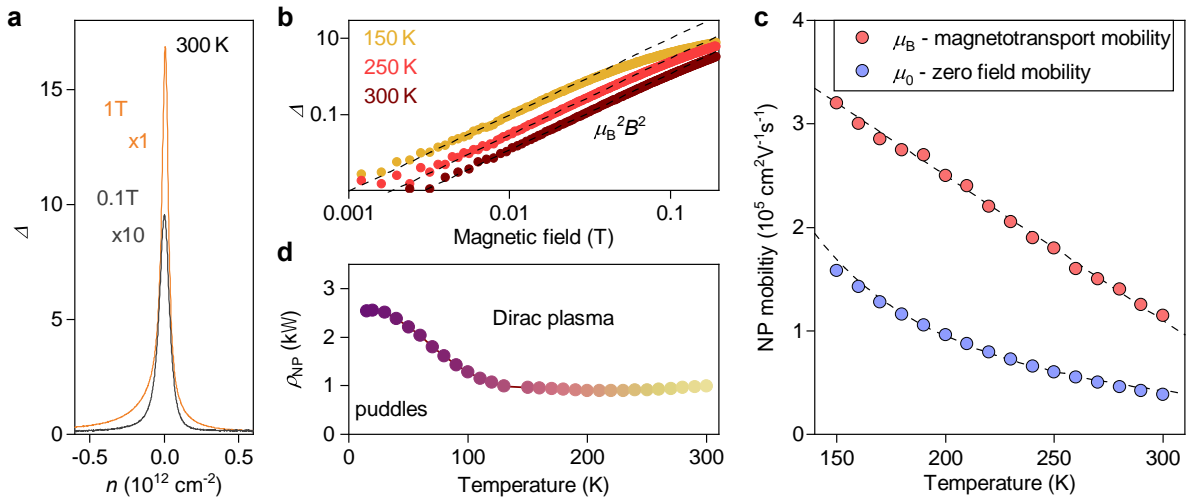

**Fig. 2| Ultrahigh mobility of the Dirac plasma. a**, $\Delta$ at the NP as a function of carrier density for characteristic $B$ at 300 K. Note 10 times different scales for the two curves. The arrow indicates the thermal carrier density. **b**, $\Delta$ at the NP plotted on the log scale. Dashed lines: parabolic fits for $B < 50$ mT. **c**, Zero-field and magnetotransport mobilities at the NP. **d**, Resistivity of MLG at the NP as a function of $T$. The shadowed region indicates the range where electron transport is affected by e-h puddles. Data in panels **c-d** are from device D1; panel **a** from D2 (also, see Extended Data Fig. 1).

It is instructive to compare the found $\mu_B$ with the zero-field mobility $\mu_0$. The latter can be evaluated using the standard Drude formula $\rho_{NP}^{-1} = 2n_{th}e\mu_0$ where $e$ is the electron charge and the factor 2 accounts for equal concentrations of electrons and holes at the NP. Fig. 2d shows that $\rho_{NP}$ quickly decreases with increasing $T$ from liquid-helium to ~100 K but, as the Dirac plasma gets established ($n_{th}$ >> residual $\delta n$), $\rho_{NP}$ becomes almost $T$ independent with a constant value of ~1 kOhm above 150 K (also, inset of Fig. 3b, Extended Data Fig. 1). The saturating behavior of $\rho_{NP}$ is attributed to the onset of the quantum-critical regime in which the scattering is dominated by the Planckian frequency, $\tau_P^{-1}$. Indeed, $\rho_{NP} \approx 1$ kOhm yields $C \approx 0.7$ close to unity, as expected[3–5,9–12]. This analysis also agrees with that of the quantum-critical behavior reported for BLG[9,10] (Methods) and conclusions about MLG from other measurements[5].

Fig. 2c shows that $\mu_0$ evolves $\propto 1/T^2$, as expected for Planckian systems with the Dirac spectrum (Methods). Surprisingly, $\mu_0$ is 2-3 times smaller than $\mu_B$. As shown in Methods, this happens because $\mu_B$ is less sensitive than $\mu_0$ to the dominating e-h scattering. Qualitatively, the difference can be understood as arising from different relative motions of electrons and holes in zero and finite $B$. In zero $B$, the electric field forces electrons and holes to move in opposite directions so that e-h collisions are efficient in impeding a current flow. In contrast, cyclotron motion causes a drift of both electrons and holes in the same direction. Therefore, e-h collisions do not affect Hall currents responsible for magnetoresistivity. This explanation is further substantiated by our measurements using screened graphene devices (encapsulated MLG with metallic gates placed at a distance of ~1–3 nm)[15]. The screening is found to suppress Coulomb scattering, which results in smaller $C$ and, therefore, higher $\mu_0$ (Extended Data Fig. 4a). However, the same screening has little effect on $\rho_{NP}(B)$ and hence $\mu_B$ (Extended Data Fig. 4b), in agreement with theory. This consideration is equally applicable for e-h plasma of massive fermions and, indeed, a similar difference between $\mu_0$ and $\mu_B$ is observed for neutral BLG (Extended Data Fig. 2). The above analysis allows us to conclude that the anomalously large MR in low $B$ arises due to ultrahigh mobility of Dirac fermions, combined with ineffectiveness of e-h scattering in suppressing Hall currents.

**Strange linear magnetoresistance in the extreme quantum limit**
In high $B$, magnetotransport in the Dirac plasma exhibits profound changes such that, above a few T, $\rho_{NP}(B)$ evolves from being parabolic into linear (Fig. 3, Extended Data Fig. 5). Slopes of this linear MR are found to be similar for all the studied devices (inset of Fig. 3b) and almost independent of $T$. The crossover between parabolic and linear dependences is marked by a flattened section on the curves which appears at $T$ below 200 K. We attribute the flattening to the onset of Landau quantization (Extended Data Fig. 6). This attribution also agrees with the fact that, at $B \approx 3\text{-}5$ T, the main cyclotron gap between the zeroth and first Landau levels (LLs) reaches ~800 K, notably exceeding the thermal smearing $k_BT$. As for the MR magnitude, $\Delta$ reaches ~$10^4$ % at 10 T and, despite the linear (slower than parabolic) dependence in quantizing fields, this is again record-high for room-$T$ experiments[32]. Comparison with multilayer and low-quality graphene (Extended Data Figs. 3&8) shows the importance of both Dirac spectrum and electronic quality for such a giant MR response. Another notable feature of magnetotransport in zeroth-LL's plasma is that $\rho_{NP}$ at a given $B$ increases with increasing $T$ (Fig. 3a, Extended Data Fig. 5). This contradicts the orthodox MR behavior observed in all other systems, which results in lower $\Delta$ at higher $T$ because of increased scattering[29] (Methods). To shed light on strange magnetotransport, we have also tested how $\rho_{NP}(B)$ is affected by proximity screening. Although the parabolic dependence of $\rho_{NP}$ in low $B$ was practically unaffected (as discussed above), the screening greatly suppressed MR in quantizing $B$ (Fig. 3b). The linear slopes of $\rho_{NP}(B)$ decrease from 5–8 k$\Omega$ T$^{-1}$ in our primary devices to 1–3 k$\Omega$ T$^{-1}$ in those with screening (inset of Fig. 3b), implying that magnetotransport on the zeroth LL depends on Coulomb interactions.

In discussing the high-$B$ behavior, we first note that the previously reported linear MR can in most cases be attributed to complex current flows that become increasingly nonuniform as $\rho_{xy} \propto B$ increases (Sections 1&2 of Methods). The involved mechanisms are based on either spatial inhomogeneity or the presence of edges. To check for possible edge effects in our case, we have studied Corbino disks fabricated from encapsulated MLG and found very similar $\rho_{NP}(B)$ (Extended Data Fig. 7). Thus, for our zeroth-LL plasma with zero $\rho_{xy}$, those extrinsic mechanisms can be ruled out (Methods). It may also be tempting to evoke Abrikosov's linear magnetoresistance[16] predicted to occur in 3D semimetals with Dirac-like spectra in the extreme quantum limit. However, the Born approximation used in the 3D model cannot be justified for 2D transport in a smooth background potential[37] because charge carriers remain localized within electron and hole puddles.

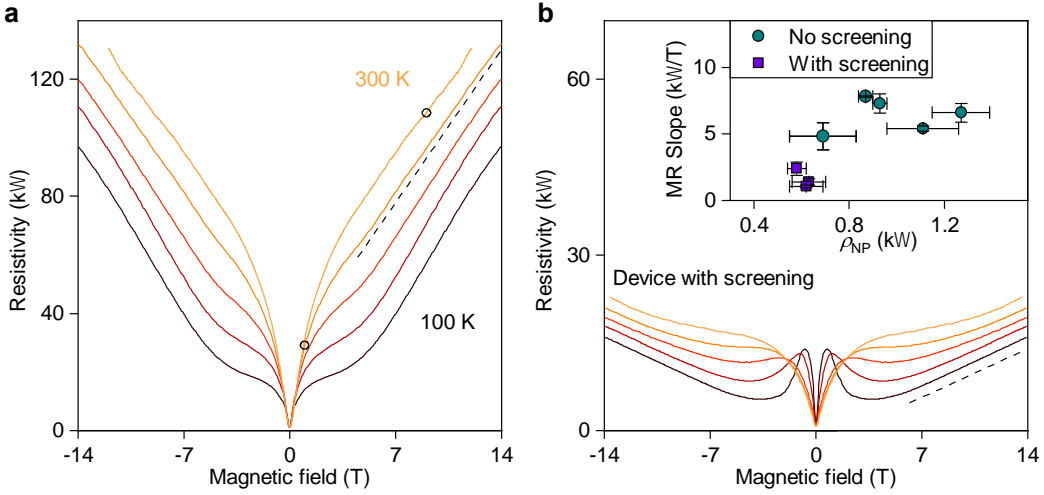

**Fig. 3| Linear magnetoresistance in quantizing fields. a**, Magnetoresistivity of the neutral Dirac plasma between 100 and 300 K in steps of 50 K. The black circles mark $B = 1$ and 9 T where $\Delta$ reaches ~2,500 and 8,600 %, respectively. The $B$ values are chosen for easier comparison with the highest MR observed previously, as summarized in ref.[32]. Dashed line: guide to the eye with a slope of 7.3 kΩ T$^{-1}$. The data are for device D1. Device D2 exhibits similar behavior (Extended Data Fig. 5). **b**, $\rho_{NP}(B)$ for the screened Dirac plasma (color coding as in **a**). Dashed line: 1.2 kΩ T$^{-1}$. Inset: linear MR's slopes as a function of zero-field $\rho_{NP}$ for three devices with proximity screening and five standard devices (Methods). Error bars: SD including small changes in the slopes with $T$ and variations observed for different sections of our multiterminal devices.

For the lack of a theory suitable to describe the observed linear MR, we employ the simple Drude model by considering cyclotron-orbit centers as quasiparticles that circle along equipotential contours and, also, diffuse between them due to electron scattering. The density of such quasiparticles is determined by LL's capacity, $n_{LL} = 2B/\phi_0 \gg n_{th}$. For charge neutrality, the Drude model yields $\rho_{NP}(B) \approx \rho\mu^2 B^2$ (Methods) where $n$ and $\mu$ in the standard expression $\rho = 1/ne\mu$ should be substituted with $n_{LL}$ and $\mu_Q$, respectively, to reflect the density and mobility for zeroth-LL's plasma. This leads to

$$\rho_{NP}(B) \approx \frac{h}{2e^2} \mu_Q B \qquad (1)$$

The linearity in $B$ arises from the fact that the $B^2$ dependence inherent for compensated semimetals is moderated by the linear increase in the carrier density on the zeroth LL. Next, to estimate $\mu_Q$, we assume that Planckian scattering moves quasiparticles by a typical distance $\ell$ between equipotentials, resulting in the diffusion coefficient $D \approx \ell^2/\tau_p = v_T^2 \tau_p$ with the corresponding thermal velocity $v_T \equiv \ell/\tau_p$. Diffusion within individual puddles leaves carriers localized inside. Only if a quasiparticle covers a distance of ~$\xi$ between neighboring puddles, those processes contribute to macroscopic currents along the electric field and, hence, global conductivity. Accordingly, the time scale relevant for electron transport on the zeroth LL is given by $\tau_{tr} \approx \xi^2/D \gg \tau_p$ and the corresponding diffusion coefficient can be written as $D_{tr} = v_T^2 \tau_{tr} \approx v_T^2 \xi^2/D = \xi^2/\tau_p$. Then, using the Einstein-Smoluchowski equation, we find the transport mobility $\mu_Q = eD_{tr}/k_B T \approx e\xi^2/k_B T \tau_p = Ce\xi^2/h$ where both $v_T$ and $\ell$ fell out from the final expression. This result suggests zero-LL's MR to be linear in $B$ and independent of $T$, as observed experimentally, and may also explain the suppression by proximity screening as smaller $C$ result in lower $\mu_Q$. Furthermore, eq. 1 can be rewritten as

$$\rho_{NP} \approx \frac{C}{4\pi} \frac{h}{e^2} \frac{\xi^2}{\ell_B^2} \qquad (2)$$

which closely resembles the result of a formal extension of Abrikosov's model into the 2D case[37]. Although the above consideration catches the main physics and qualitatively agrees with our observations, further work is required to develop a microscopic theory of magnetotransport in the 2D Boltzmann plasma at the zeroth LL.

**Outlook**

The Dirac plasma in graphene exhibits the highest MR observed above liquid-nitrogen $T$ in both low and high fields. In low $B$, only ferromagnetic devices employing spin tunneling[38] or the use of four-probe geometry[26,33] allow stronger electronic response to magnetic fields. In contrast to the latter phenomena, the giant MR of graphene stems from its magnetoresistivity $\rho_{xx}(B)$. In quantizing fields, graphene experiences a system transformation becoming an e-h plasma residing on the zeroth LL. Our observations are also relevant to the physics of strange metals that exhibit Planckian scattering. Strange metals display the renowned linear $T$ dependence of their resistivity, in obvious contrast to our case. However, this difference arises only because strange metals have a fixed carrier density whereas the carrier density and effective mass in the Dirac plasma increase with $T$, leading to the constant $\rho_{NP}$. Moreover, strange metals also exhibit linear MR that is weakly $T$ dependent. This MR remains unexplained, although a recent ansatz[13,14] suggests that, in Planckian systems, $\tau^{-1}$ should be defined by the largest relevant energy scale, either $k_B T$ or some magnetic-field-induced energy $\propto B$. The ansatz does not seem work for the Dirac plasma because the only relevant and sufficiently large magnetic energy is the cyclotron gap. It evolves as $B^{1/2}$ rather than linearly in $B$. Notwithstanding any differences, Planckian systems in high fields remain poorly studied, and graphene's Dirac plasma offers a model system to understand the relevant physics. The possibility to modify magnetotransport by tuning electron-electron interactions using proximity screening is especially appealing in this context.

## METHODS

### 1. Brief history of linear magnetoresistance

Studies of the electrical response of metals to magnetic fields go back to experiments by Lord Kelvin and Edwin Hall over one-and-a-half century ago[39,40]. Although the subject continued to attract sporadic attention during the following decades (see, e.g., ref.[41]), the first systematic study of magnetoresistance phenomena is usually credited to Pyotr Kapitsa. In 1928-29, he reported high-field studies of MR in 37 different materials[42,43]. This research brought up two major findings. First, some materials (e.g., Bi, As, Sb and graphite) were found to exhibit magnetoresistance exceeding 100% in 30 T at room $T$, much higher than the others in that study. So large MR could not be explained by contemporary theories. Second, despite different absolute values of MR, all the studied materials followed a universal $B$ dependence. In small fields, it was always parabolic, in agreement with the already accepted understanding that cyclotron motion of current-carrying electrons should bend their trajectories and, hence, increase resistivity. However, in fields above several T, MR was found to increase linearly, which was unexpected.

The first puzzle of large MR values was solved relatively quickly, thanks to the development of the band theory. Most of the materials exhibiting large room-$T$ MR in Kapitsa's experiments appeared to be semimetals so that the electric current was carried by both electrons and holes. It is now well understood that the reduced Hall effect in this case leads to $\rho_{xx}$ evolving in high fields approximately as $1/\sigma_{xx}$, in contrast to the case of one type of charge carriers where $\rho_{xx} \approx \sigma_{xx}\rho_{xy}^2$ (Section 3 below). The second puzzle of linear magnetoresistance has attracted numerous theories and explanations. Generally, there are several mechanisms that can cause linear MR and, even today, its observation often leads to controversies because it is difficult to pinpoint the exact origin.

One of the first mechanisms causing linear magnetoresistance was proposed by Lifshitz and Peschanskii[44]. In 1959, they considered magnetotransport in polycrystalline metals with open Fermi surfaces. For certain orientations of the magnetic field with respect to crystallographic axes, such metals flaunt open cyclotron orbits that result in non-saturating MR $\propto B^2$ (refs.[45,46]). However, this quadratic behavior occurs only within a narrow interval of angles which decreases as $\propto B^{-1}$. For the other angles, cyclotron orbits remain closed, and MR attributable to them saturates in high $B$. Averaging over all angles for polycrystalline samples resulted in linear MR, and this result helped to explain many - but not all - observations in the literature. Those ideas were further developed by Dreizin and Dykhne[47] who obtained MR $\propto B^{4/3}$ and $B^{2/3}$, depending on whether a metal with an open Fermi surface was compensated or not, respectively. Moreover, the authors presented a magnetotransport theory not only for polycrystalline but also inhomogeneous conducting media. Depending on the Fermi surface and compensation between charge carriers, various powers of $B$ could be obtained including, for example, linear MR in compensated semimetals with 2D disorder[47].

The magnetoresistance theory relying on materials' inhomogeneity was expanded both theoretically and experimentally in the 1970s and '80s. It was shown that macroscopic strain[48], voids[49–52] and thickness variations[53,54] could lead to linear MR in high $B$ ($\mu B \gg 1$). The next step was taken in 2003 by Parish and Littlewood who considered the case of very strong inhomogeneity that could not be described by the earlier theories[55]. Using a random 2D resistance network they obtained linear MR that starts from small magnetic fields ($\mu B < 1$) and could explain the behavior observed in some disordered semiconductors[55]. The fundamental reason for MR in all the cases involving inhomogeneous media is the following. In the presence of regions with different magnetotransport coefficients, the arising Hall voltages (large for $\mu B \gg 1$) necessitate substantial changes in the electric current distribution to satisfy boundary conditions at interfaces between different regions. As a result, the electric current becomes increasingly inhomogeneous, being squeezed into narrow streams near the interfaces. This current inhomogeneity increases the effective resistance of the medium[53,54].

A different mechanism was suggested by Abrikosov[16,56,57]. He pointed out that some materials exhibiting linear MR were neither polycrystalline nor inhomogeneous but single crystals with closed Fermi surfaces including graphite, bismuth and other materials[58,59]. To explain those observations, Abrikosov considered a Weyl (3D Dirac-like) spectrum so that, in quantizing $B$, all charge carriers collapsed onto the lowest (zero) Landau level. Assuming a scattering potential caused by screened charged impurities, linear MR was predicted in this case. Because of the essential role played by Landau quantization, the effect was called quantum linear magnetoresistance[16,56,57]. The Abrikosov mechanism attracted considerable interest and was invoked as an explanation for many experiments[60,61], even though the concerned materials often poorly matched the assumptions required by the theory (including being 2D rather than 3D systems). Unfortunately, Abrikosov provided no explanation for the physics behind his theory and, only recently[37], it has been shown that his analysis is equivalent to calculations of diffusion of cyclotron-orbit centers in an electrostatic potential. This conceptual overlap requires mentioning of the earlier theories by Kubo and Ando for diffusion of cyclotron-orbit centers[62,63]. Furthermore, within the self-consistent Bohr approximation, linear MR was shown to appear in the 2D case for strongly screened charged impurities whereas, for short-range scattering, MR becomes sublinear[64]. The formal extension of Abrikosov's theory into 2D also leads to linear MR[37].

Finally, two other mechanisms that result in linear MR have to be mentioned. First, Alekseev with colleagues showed that e-h annihilation at the edges of 2D semimetals could lead to linear magnetoresistance[65,66]. This mechanism can be ruled in or out by comparing magnetotransport in Hall-bar and Corbino-disk devices, as done in our work. Second, so-called strange metals often exhibit resistivity that increases linearly not only with temperature but also with magnetic field[13,14]. While such linear MR does not follow from the so-called holographic approach[11], it was suggested[13,14,67] that the quantum critical scattering rate $\tau^{-1} \approx E/h$ could be controlled by the maximum relevant energy $E$ in the uncertainty equation, that is, by either $k_B T$ or $\mu_B B$ where $\mu_B$ is the Bohr magneton. This could then explain both ($T$ and $B$) linear dependences in strange metals. It is also worth mentioning that linear MR was recently reported in two other 2D strongly-interacting systems, namely, twisted WSe$_2$[68] and magic-angle graphene[69]. It was suggested that the MR had the same origin as in strange metals.

## 2. Earlier studies of magnetoresistance in graphene and Dirac-type materials
Over the last decade, there were numerous studies of magnetotransport using newly available materials such as graphene (see, e.g., refs.[30–34,61,70–74]), topological insulators (see, e.g., refs.[75–77]) and high-mobility

Dirac/Weyl semimetals (see, e.g., refs.[17–25]). Graphene attracted particular attention as a promising material for magnetic-field sensors due to its high $\mu$ at room $T$. The first generation of graphene devices (graphene placed on oxidized silicon and so-called epitaxial graphene) exhibited relatively low $\mu$, and their MR was also relatively low, reaching only ~100% in fields above 10 T (refs.[30–32,61,70–72]; Section 13 below). The MR typically originated from charge inhomogeneity and other disorder, although some reports suggested[61] the observation of Abrikosov's linear MR in doped multilayer graphene. Later research ruled out this explanation, arguing that the observed linear MR originated from a polycrystalline disorder[31,55].

The next generation of graphene devices using encapsulation with hBN exhibited exceptional electronic quality[2,78]. So far, magnetotransport in graphene-on-hBN devices was studied at elevated $T$ only for few-layer graphene[34] and MLG away from the NP[33]. Few- and multi- layer graphene (graphite) exhibit a relatively low $\mu$ at elevated temperatures. This results in small quadratic MR in low $B$ and also limits MR in high fields[32,34], in agreement with our results in Section 6 below. Doped high-$\mu$ graphene exhibits saturating magnetoresistivity and its magnitude is small, as expected. Note however that, if one uses a geometry that instigates a nonuniform current flow, it is possible to enhance the apparent MR in 4-probe measurements, for example using the so-called 'extraordinary magnetoresistance' configuration[26,33]. In such a geometry, the central part of a MLG device is replaced with a highly conducting metal (e.g., gold). In zero $B$, the current mainly flows through the metal, despite being injected into graphene. Magnetic field curves the current trajectories and forces charge carriers to move through graphene, which is much more resistive than gold films. Accordingly, the apparent 4-probe magnetoresistance could reach extraordinary values of $\sim 10^7$ % at 9 T and room $T$. This is comparable to typical changes in Hall voltage that also requires a 4-probe geometry. Note that this extraordinary MR is not an intrinsic property of a material and, accordingly, translates into only modest changes for any 2-probe measurements. Until now, no studies of magnetotransport at elevated $T$ have been reported for charge-neutral monolayer graphene with high $\mu$.

For completeness, let us mention extensive magnetotransport studies of 3D counterparts of graphene, which are different topological insulators, Dirac/Weyl semimetals and other clean semimetals like $WTe_2$ (also suggested to be a Weyl semimetal[79]). Many of them showed huge MR, which in some cases exceeded $10^6$ % at liquid-helium temperatures[17–22,25]. Such colossal values were attributed to high mobility of charge carriers in these materials ($\mu$ reaching above $10^6$ cm$^2$ V$^{-1}$ s$^{-1}$ at 4 K, similar to encapsulated graphene). However, the mobility rapidly decayed with increasing $T$, which resulted only in a tiny low-$B$ MR at elevated $T$. This is not the case for monolayer graphene that exhibits high $\mu$ at room $T$ even at the NP, which results in the colossal quadratic MR in low $B$, as reported in this work.

### 3. Drude model for charge-neutral graphene

To evaluate magnetotransport properties of our devices, we have used the standard two-carrier model for electrons and holes, which allows the longitudinal and Hall conductivities to be written as[80]

$$\sigma_{xx}(B) = \frac{en_e\mu_e}{1+(\mu_e B)^2} + \frac{en_h\mu_h}{1+(\mu_h B)^2} \tag{S1}$$

$$\sigma_{xy}(B) = \frac{en_h\mu_h^2 B}{1+(\mu_h B)^2} - \frac{en_e\mu_e^2 B}{1+(\mu_e B)^2} \tag{S2}$$

where $n_{e(h)}$ is the carrier density of electrons (holes) and $\mu_{e(h)}$ is the corresponding mobility. The relative magnetoresistivity is defined as

$$\Delta = [\rho_{xx}(B) - \rho_{xx}(0)]/\rho_{xx}(0) \tag{S3}$$

where $\rho_{xx}(B) = \frac{\sigma_{xx}(B)}{\sigma_{xy}^2(B)+\sigma_{xx}^2(B)}$. For the case of a compensated semimetal with $n_e = n_h$ and equal mobilities for electrons and holes ($\mu_e = \mu_h = \mu$), the above equations yield

$$\Delta = \mu^2 B^2 \tag{S4}$$

This expression was used in this work to extract the magnetotransport mobility $\mu_B$ from parabolic dependences of $\rho_{NP}(B)$ in small $B$.

Our analysis of the $\rho_{xx}(n)$-peak broadening and the zero-field mobility $\mu_0$ (see the main text) have relied on theoretical expressions for the density of thermally excited electrons at the NP, $n_{th}$. For monolayer graphene, this electron density is given by

$$n_{th} = \int_0^{+\infty} f(E)\, DOS\, dE = \int_0^{+\infty} \frac{1}{\exp\left(\frac{E}{k_B T}\right)+1} \frac{2E}{\pi \hbar^2 v_F^2} dE = \frac{2(k_B T)^2}{\pi \hbar^2 v_F^2} \int_0^{+\infty} \frac{x}{\exp(x)+1} dx = \frac{2\pi^3}{3} \frac{(k_B T)^2}{h^2 v_F^2} \tag{S5}$$

where $\hbar = h/2\pi$ is the reduced Planck constant. Holes are excited with the same density. Thermally-excited Dirac fermions with a typical energy $k_B T$ can be assigned with the effective mass $m^*$ that is also $T$ dependent

$$m^* = \pi^2 k_B T/(6\ln 2) v_F^2 \tag{S6}$$

This expression can be obtained from the Boltzmann equations calculating the response of charge-neutral graphene to electric field and enforcing the resulting conductivity into a Drude-like form. Note that the above mass is proportional to the typical energy (thermal energy $k_B T$ of electrons and holes in the Dirac plasma) divided by their velocity squared, as expected for ultra-relativistic particles.

Using the same approach for bilayer graphene, we obtain its density of thermally exited electrons

$$n_{th} = \frac{2\ln(2)}{\pi \hbar^2} m^* k_B T \tag{S7}$$

The above expressions for $n_{th}$ and $m^*$ have been used to evaluate conductivities of both charge-neutral MLG and BLG based on the standard Drude-like expression

$$\rho_{NP}^{-1} = 2 n_{th} e^2 \tau / m^* \tag{S8}$$

where $\tau$ is the scattering time, and the factor of 2 accounts for equal densities of thermally-excited electrons and holes.

### 4. Additional examples of magnetotransport measurements for monolayer graphene

Several (> 10) monolayer devices (Hall bars and Corbino disks) were studied during the course of this work. To indicate variations in their magnetotransport behavior, below we present measurements for another Hall bar (device D2) exhibiting notably higher remnant $\delta n$ at low $T$. Its resistivity $\rho_{NP}(T)$ at the NP is plotted in Extended Data Fig. 1a. Similar to device D1 (Fig. 2c of the main text), $\rho_{NP}$ of device D2 decreases with $T$ and saturates above 200 K. In this device, the saturation occurs at higher $T$ than in device D1 because of stronger inhomogeneity (cf. Fig. 1d of the main text and Extended Data Fig. 1b). Despite an-order-of-magnitude different inhomogeneities, both devices exhibit practically the same saturation value, $\rho_{NP} \approx 1$ kOhm. The same was valid for the other MLG devices.

As discussed in the main text, we attribute the $T$ independent $\rho_{NP}$ in MLG to the entry of the Dirac plasma into the quantum-critical regime[3–5,9–12]. In this regime, the electron scattering time is determined by Heisenberg's uncertainty principle, $\tau_p^{-1} = C \frac{k_B T}{h}$ where $C$ is the interaction constant of about unity and depends on screening[3,4,11,12]. By plugging this scattering rate into eq. S8 and using the effective mass from eq. S6 and the carrier density given by eq. S5, we obtain the quantum-critical resistivity

$$\rho_{NP} = C(h/e^2)/8\pi\ln 2 \tag{S9}$$

which is independent of $T$. The observed $\rho_{NP} \approx 1$ kOhm yields the interaction constant $C \approx 0.7$, close to unity, as expected for Planckian-limit scattering[3–5,9–12].

As for the MR behavior of device D2, Extended Data Fig. 1c shows that $\rho_{NP}$ is parabolic in low $B$, similar to the case of device D1 in Fig. 1 of the main text. The absolute value of $\Delta$ for device D2 is also similar, albeit slightly smaller, reaching 90% at 0.1 T and room $T$. The 20% reduction can be attributed to the lower electronic quality and homogeneity of device D2. Extended Data Fig. 1d plots zero-field and magnetotransport mobilities for device D2, which were extracted using the same approach as described in the main text. Both mobilities are slightly lower than those in Fig. 2 of the main text. Nonetheless, at room temperature, $\mu_B$ in device D2 still exceeds 100,000 cm$^2$ V$^{-1}$ s$^{-1}$. Overall, the results presented in Extended Data Fig. 1 corroborate our conclusions that the Dirac plasma flaunts exceptionally high carrier mobility at elevated $T$, with no analogues among compensated metallic systems. The figure also reiterates the considerable differences between $\mu_B$ and $\mu_0$, which were discussed in the main text and explained in Section 7 below.

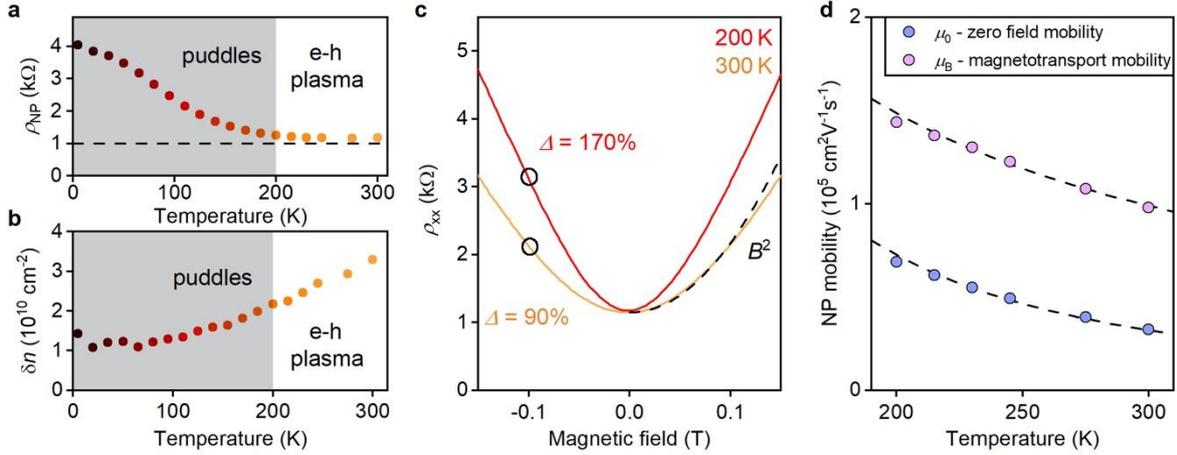

**Extended Data Fig. 1| Magnetoresistance behavior for another MLG device. a**, Its zero-$B$ resistivity at the NP as a function of $T$. **b**, $\delta n$ as a function of $T$. The shadowed areas in **a** and **b** indicate the range in which the thermally-excited density $n_{th}$ is less than the remnant inhomogeneity. **c**, Low-$B$ resistivity for two characteristic $T$. Dashed curve: parabolic dependence. The black circles indicate $\Delta$ in the characteristic field of 0.1 T. **d**, $\mu_B$ and $\mu_0$ for this device as a function of $T$. Dashed curves: guides to the eye.

## 5. Electron-hole plasma in bilayer graphene

To emphasize how unique the Dirac plasma in MLG is, let us compare its magnetotransport properties with those of the closest electronic analogue, an e-h plasma at the NP in bilayer graphene. To this end, we fabricated and studied BLG devices that were also encapsulated in hBN to achieve high $\mu$. They were double-gated and shaped into the standard Hall bars. At liquid-helium $T$ and away from the NP, the devices exhibited ballistic transport across their entire widths of ~10 μm. This was observed directly using bend resistance measurements[81]. The double-gating was required to tune the carrier density to the NP while maintaining zero bias between the two graphene layers. The latter ensured that no gap opened at the NP[82], which otherwise would complicate the comparison[10].

Typical behavior of BLG's resistivity in zero $B$ is shown in Extended Data Fig. 2a. Similar to MLG (Fig. 1b of the main text; Extended Data Fig. 1a), $\rho_{NP}(B=0)$ of charge-neutral BLG reaches a few kOhm at liquid-helium $T$, but rapidly decreases to ~1 kOhm at higher $T$ and becomes $T$ independent above 50 K (Extended Data Fig. 2b). Such behavior of high-quality BLG has already been reported recently, and constant $\rho_{NP}$ was attributed to the e-h plasma entering the quantum-critical regime[9,10]. Indeed, plugging the quantum critical scattering rate $\tau_p^{-1} = C\frac{k_B T}{\hbar}$ into eq. S8 and using the thermally excited density from eq. S7, we obtain the resistivity for the e-h plasma in BLG

$$\rho_{NP} = C(h/e^2)/16\pi\ln2 \tag{S10}$$

The $T$ independent value of $\rho_{NP}$ stems from the fact that both $n_{th}$ and scattering frequency $\tau_p^{-1}$ evolve linearly with $T$. Eq. S10 differs from eq. S9 for MLG only by a factor of 2. From the data of Extended Data Fig. 2, we obtain $C \approx 1.4$, close to unity as expected and in agreement with the previous reports[9,10]. This value is twice larger than $C$ for the Dirac plasma in MLG. We are unaware of any theory that would allow quantitative comparison between $C$ in the two graphene systems. Nonetheless, the smaller value of the interaction constant in MLG as compared to BLG could probably be understood as due to the lower density of states in the Dirac spectrum.

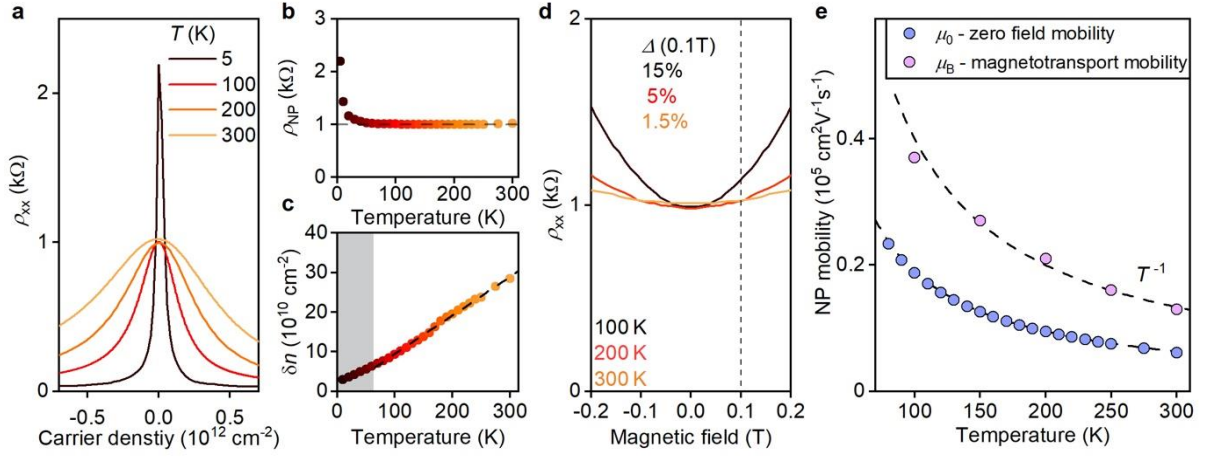

**Extended Data Fig. 2| Magnetoresistance and mobility of e-h plasma in bilayer graphene. a**, Its zero-$B$ resistivity near the NP as a function of gate-induced carrier density. **b**, Resistivity of BLG at the NP as a function of $T$. **c**, Thermal smearing $\delta n$ was extracted using the same approach as described in the main text. **d**, Magnetoresistivity at the NP in low $B$. **e**, Carrier mobility at the NP as a function of $T$. $\mu_B$ and $\mu_0$ were extracted from magnetoresistance and zero-field measurements, respectively. Dashed lines: $1/T$ dependences. All the measurements were carried out at zero displacement.

Additionally, we analyzed $\delta n(T)$ for our BLG devices using the same approach as described for MLG in the main text. Above 50 K, $\delta n$ in Extended Data Fig. 2c exceeds the remnant charge inhomogeneity (in the limit of low $T$) by a few times, which ensures that the smearing of the peak in $\rho_{xx}$ at $T > 100$ K was dominated by e-h excitations. Extended Data Fig. 2c also shows that $\delta n$ in BLG increased linearly with $T$, in agreement with eq. S7 and qualitatively different from the quadratic behavior of $\delta n(T)$ in MLG (eq. S5; Fig. 1d of the main text). Using the usually assumed value $m^* \approx 0.03\, m_e$ for BLG ($m_e$ is the free electron mass), we find $\delta n \approx 0.5 n_{th}$, similar to the case of MLG as discussed in the main text.

The response of BLG's e-h plasma to small $B$ is shown in Extended Data Fig. 2d. Similar to the case of MLG, $\Delta$ evolves $\propto B^2$ but its absolute value is two orders of magnitude smaller than that in MLG, reaching only 1.5% at 0.1 T at room $T$. For completeness, we have evaluated the mobilities for the compensated e-h plasma in BLG, using the same approach as in the main text. Both magnetotransport and zero-field mobilities ($\mu_B$ and $\mu_0$, respectively) are plotted in Extended Data Fig. 2e. They are found to be an order of magnitude lower than those for the Dirac plasma, which is the underlying reason behind the two-orders-of-magnitude smaller low-$B$ magnetoresistance in BLG as compared to MLG ($\Delta \propto \mu^2$). Note that $\mu_0$ for BLG is approximately twice lower than $\mu_B$ (Extended Data Fig. 2e), similar to the case of MLG in Fig. 2c of the main text. The difference between $\mu_0$ and $\mu_B$ is again attributed to electrons and holes moving against and along each other for longitudinal and Hall flows, respectively, as discussed in the main text and detailed in Section 7 below.

Our experiments show that charge carriers in the Dirac plasma are several times more mobile than electrons and holes at the NP in BLG. The reason for the exceptionally high mobility in the Dirac plasma is twofold. First, the scattering rate $\tau_p^{-1} \propto C$ is approximately twice lower in MLG than BLG, as discussed above. Second, the effective mass for Dirac fermions at room $T$ can be estimated from eq. S6 as $m^* \approx 0.01 m_e$, which is 3 times lower than the effective mass of charge carriers in BLG. Taken together, this suggests that the zero-field mobility $\mu_0 = e\tau/m^*$ for the Dirac plasma should be a factor of 6 higher than that for BLG's e-h plasma, in qualitative agreement with the experiment (cf. Extended Data Fig. 1d and 2e).

## 6. Magnetotransport in multilayer graphene

Another electronic system with high-mobility charge carriers at room $T$ is multilayer graphene (thin films of graphite). The material is an intrinsic semimetal with electrons and holes being present in approximately same concentrations[83]. It is instructive to compare the magnetotransport properties of this nearly compensated semimetal with those of the Dirac plasma.

Our graphite devices were several nm thick (10–20 graphene layers) and shaped into Hall bars. To preserve the high electronic quality, the multilayer films were again encapsulated with hBN. Measurements for one of the devices are shown in Extended Data Fig. 3. Graphite's magnetoresistivity was found to increase quadratically in fields below 1 T. At room $T$, $\Delta$ was ~1.4% at 0.1 T, similar to the case of BLG and two orders of magnitude smaller than MR of the Dirac plasma. Above 1 T, graphite exhibited notable deviations from the parabolic dependence bending towards a lower power and becoming practically linear in $B$ at low $T$ and above a few T. Room-temperature $\Delta$ reaches 80% and 3,500% at 1 and 9 T, respectively, in agreement with the previous report for few-layer graphene[34]. Although MLG exhibits a few times larger $\Delta$ in high $B$, it is possible that the linear MR in graphite (first reported a century ago[42,43] and still not fully understood; see Section 1) has the same origin as in monolayer graphene. This possibility requires further investigation because graphite's electronic spectrum is complicated and, also, strongly evolves with magnetic field[83].

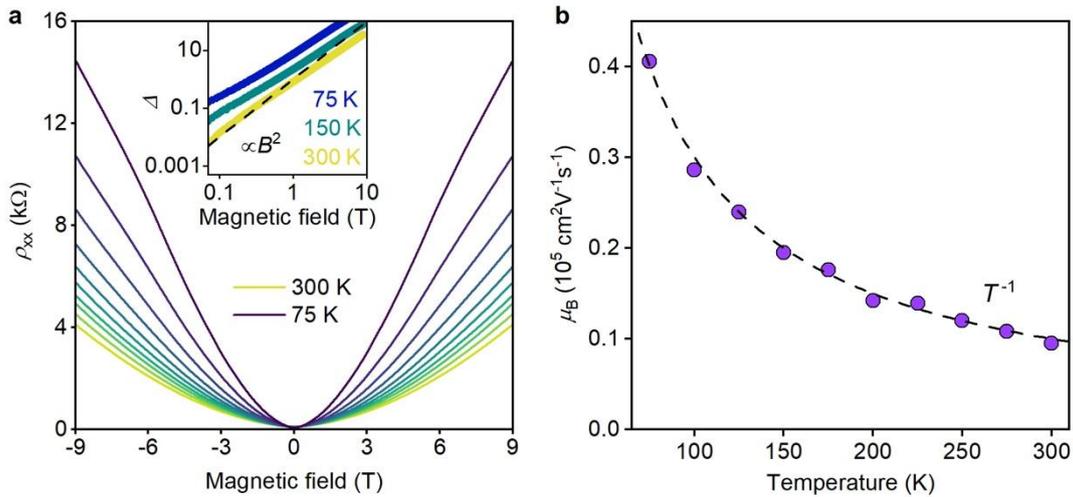

**Extended Data Fig. 3| Magnetoresistivity of multilayer graphene. a,** $\rho_{xx}(B)$ for a 3.5 nm thick graphite film (~10 graphene layers) measured between 75 and 300 K in steps of 25 K. Inset: Log-log plot of the magnetoresistance at three characteristic $T$. Dashed line: quadratic dependence. **b,** Magnetotransport mobility for graphite evaluated using parabolic fits of $\rho_{xx}(B)$ over an interval of ±0.5 T. Dashed curve: $1/T$ dependence. We measured several graphite devices, and all exhibited very similar MR behavior. It changed little if additional carriers were induced near the surface by gate voltage[83].

To evaluate the magnetotransport mobility $\mu_B$ in graphite, we used the same approach as for mono- and bi-layer graphene. The results are plotted in Extended Data Fig. 3b. At room $T$, $\mu_B$ for the e-h system in graphite was found to be ~10,000 cm$^2$ V$^{-1}$ s$^{-1}$, that is, a factor of >10 lower than that for the Dirac plasma in MLG (Fig. 2b of the main text) but close to $\mu_B$ found for the e-h plasma in BLG (Extended Data Fig. 2e). This is perhaps not surprising as electronically graphite is often considered as a stack of graphene bilayers. The provided comparison of graphene with its bilayers and multilayers highlights the unique nature of the Dirac plasma and its anomalously high mobility that results in the giant MR response, especially in low $B$. Note that $\mu_B$ for multilayer graphene can be extracted more accurately, using both Hall and longitudinal measurements, which does not require the used assumption of e-h symmetry at the NP. The latter analysis[83] yields practically the same $\mu_B$ as our intentionally simplified approach.

## 7. Difference between zero-field and magnetotransport mobilities

Magnetotransport in graphene's Dirac plasma was first analyzed by Muller and Sachdev[84] and later by Narozhny with colleagues[85,86]. Below we provide analogous calculations, for completeness and to simplify our evaluation of the magnetoresistivity observed experimentally.

In the presence of electric $E$ and magnetic $B$ fields, the Boltzmann equations for electrons and holes at the NP can be written as

$$\begin{cases} -\frac{e}{m^*}(E + u_e \times B) = \frac{1}{2}\frac{u_e - u_h}{\tau_{eh}} + \frac{u_e}{\tau} \\ \frac{e}{m^*}(E + u_h \times B) = -\frac{1}{2}\frac{u_e - u_h}{\tau_{eh}} + \frac{u_h}{\tau} \end{cases} \quad (S11)$$

where $u_{e/h}$ are the drift velocities of electrons and holes, respectively, $\tau_{eh}$ is the e-h scattering time, and $\tau$ is the electron-impurity and/or electron-phonon scattering times. The effective mass $m^*$ for the Dirac plasma is given by eq. S6.

Taking the sum and difference between the top and bottom expressions in eq. S11, we obtain

$$\begin{cases} -\frac{e}{m^*}(u_e - u_h) \times B = \frac{u_e + u_h}{\tau} \\ -\frac{e}{m^*}[2E + (u_e + u_h) \times B] = \frac{u_e - u_h}{\tau_0} \end{cases} \quad (S12)$$

where $\tau_0^{-1} = \tau_{eh}^{-1} + \tau^{-1}$ is the total scattering rate. Plugging $u_e + u_h$ obtained from the top expression of eq. S12 into the left-hand side of the bottom one, we obtain

$$u_e - u_h = -\frac{2\mu_0}{1+\mu_B^2 B^2} E \quad (S13)$$

where $\mu_B^2 = \frac{e^2 \tau_0 \tau}{m^{*2}}$ and $\mu_0 = \frac{e\tau_0}{m^*}$. As shown below, these coefficients determine the magnetotransport and zero-field mobilities. If eq. S13 is placed into the left-hand side of the first line of eq. S12, this leads to

$$u_e + u_h = \frac{2\mu_B^2 B}{1+\mu_B^2 B^2} z \times E \quad (S14)$$

where $z$ is the unit vector in the direction of magnetic field. Combining eqs. S13 and S14 allows us to find

$$u_{e/h} = \mp \frac{\mu_0}{1+\mu_B^2 B^2} E - \frac{\mu_B^2 B}{1+\mu_B^2 B^2} z \times E \quad (S15)$$

Eq. S15 yields $\sigma_{xx} = (n+p)e \frac{\mu_0}{1+\mu_B^2 B^2} = 2n_{th} e \frac{\mu_0}{1+\mu_B^2 B^2}$ where $n$ and $p$ are the densities of thermally excited electrons and holes, respectively ($n = p = n_{th}$). To obtain $\rho_{xx}(B)$ at the NP, we take into account that for a compensated e-h plasma $\sigma_{xy} = 0$ and $\rho_{xx} = 1/\sigma_{xx}$, which leads to

$$\rho_{NP}(B) = \frac{1}{2n_{th} e \mu_0} + \frac{1}{2n_{th} e \mu_0}\mu_B^2 B^2 \quad (S16)$$

The first term defines the zero-$B$ resistivity of the Dirac plasma and, as expected, depends on the total scattering rate $1/\tau_0$. On the other hand, the second term is proportional to $\mu_B^2/\mu_0 = e\tau/m^*$, that is, the absolute value of magnetoresistance $\rho_{xx}(B) - \rho_{xx}(0)$ is independent of e-h collisions and depends only on impurity and/or phonon scattering.

As for relative MR, we obtain

$$\Delta = \frac{\rho_{xx}(B) - \rho_{xx}(0)}{\rho_{xx}(0)} = \mu_B^2 B^2 \quad (S17)$$

The above analysis suggests different zero-field and magnetotransport mobilities, and their ratio is given by

$$\mu_B/\mu_0 = \sqrt{\tau/\tau_0} = \sqrt{1 + \tau/\tau_{eh}} > 1 \tag{S18}$$

Our experiments found typical $\mu_B/\mu_0$ of ~3, in agreement with the expectation that e-h scattering in the Dirac plasma should be the dominant scattering mechanism at room $T$.

## 8. Effect of proximity screening on mobility and magnetoresistance

The observed difference between mobilities extracted from zero-field and magnetotransport measurements implies that $\mu_0$ and $\mu_B$ should be affected differently by screening. The latter mobility should be less sensitive to screening because e-h scattering does not contribute to Hall currents, as discussed above.

We have verified these expectations using MLG devices with proximity screening[15]. Such devices were previously studied in the doped regime where electron scattering was found to be notably reduced by the screening[15]. Electron-hole interactions in charge-neutral graphene can also be expected to be modified by such proximity screening. We studied three MLG devices in which the graphite gate served as a metallic screening plate and was separated from graphene by a thin hBN layer (thicknesses of ~0.9, 1.2 and 2.4 nm; inset of Extended Data Fig. 4a). In the particular case of the 2.4 nm device shown in Extended Data Fig. 4, we have found the screening to reduce $\rho_{NP}$ by a factor of ~2 below 250 K as compared to similar quality MLG devices without screening. The reduction in $\rho_{NP}$ yields a smaller interaction constant ($C \approx 0.4$) and translates into higher $\mu_0$. Note that the difference between $\rho_{NP}$ observed for screened and unscreened devices reduces at higher $T$ (Extended Data Fig. 4a). This can be attributed to the fact that the screening is sensitive to the average separation between charge carriers, which is $\propto n^{-1/2}$. As the density of thermally excited carriers increases with $T$, the screening efficiency is reduced[15].

The influence of proximity screening on magnetotransport in the Dirac plasma is found to be notably different from the case of zero $B$. Extended Data Fig. 4b shows that changes in $\rho_{NP}$ as a function of $B$ remained practically the same for devices with and without screening. This agrees with the results of Section 7, which predict that changes in $\rho_{NP}(B)$ should be insensitive to e-h scattering and, therefore, unaffected by proximity screening, in contrast to $\rho_{NP}(B=0)$ that is dominated by this scattering mechanism.

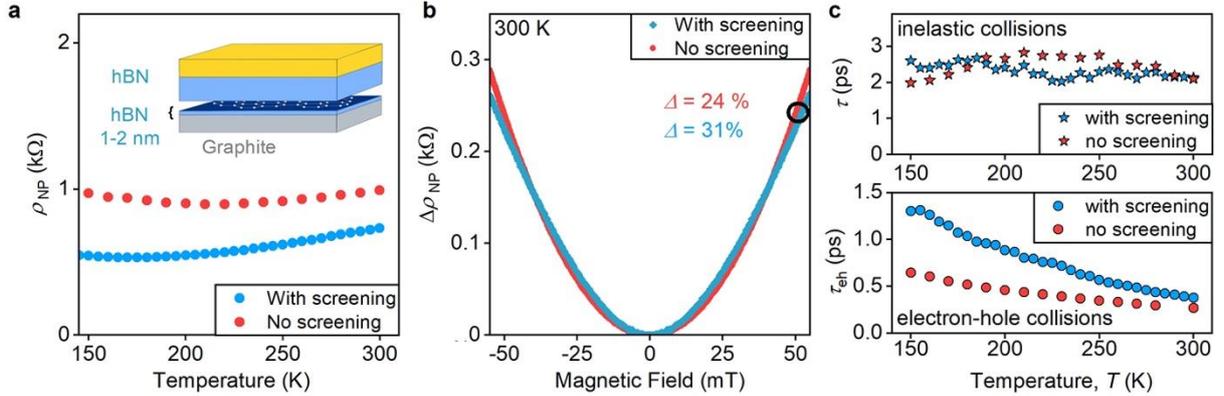

**Extended Data Fig. 4| Influence of proximity screening. a**, Resistivity of charge-neutral MLG with and without screening. Inset: schematics of our devices where the proximity screening is provided by a bottom graphite gate. **b**, Changes in graphene's resistivity in small $B$ for devices with and without proximity screening. The black circle marks $\Delta$ at 50 mT (color-coded). **c**, Inelastic and electron-hole scattering times (top and bottom panels, respectively) for devices with and without proximity screening.

For quantitative analysis of the observed screening effects, we have extracted electron-hole and electron-impurity (inelastic) scattering times ($\tau_{eh}$ and $\tau$, respectively) for the devices with and without proximity screening. To this end, we used the fact that the first (zero-$B$) term in eq. S16 depends on both $\tau_{eh}$ and $\tau$ whereas the second term is determined only by $\tau$. The results are plotted in Extended Data Fig. 4c. Both screened and unscreened devices exhibit similar $\tau_{in}$ that is several times longer than $\tau_{eh}$. As expected, the proximity screening significantly suppresses electron interactions so that, at ~150 K, $\tau_{eh}$ is twice longer in the devices with proximity screening than for the standard encapsulated graphene. The difference is reduced at higher $T$, with possible reasons for this being mentioned earlier in this section.

## 9. Linear magnetoresistivity in high fields

As discussed in the main text, the parabolic MR is observed only in small magnetic fields up to ~0.1T. In higher $B$, a linear magnetoresistance behavior emerges. We observed the linear dependence over a wide range of magnetic fields up to 18 T, highest $B$ available in our experiments. This is shown in Extended Data Fig. 5 for device D2. Again, the slope of $\rho_{NP}(B)$ depends weakly on $T$, and its absolute value is close to that exhibited by device D1 (within 20%), as shown in Fig. 3a of the main text. Overall, the described high-$B$ behavior was very similar for all five such MLG devices that we studied (inset of Fig. 3b of the main text). Note that the absence of $T$ dependence for high-$B$ MR indicates that many-body gaps caused by lifting of spin and valley degeneracies play little role within the discussed range of $T$ and $B$. Otherwise, the gaps' smearing should have led to a strong $T$ dependence.

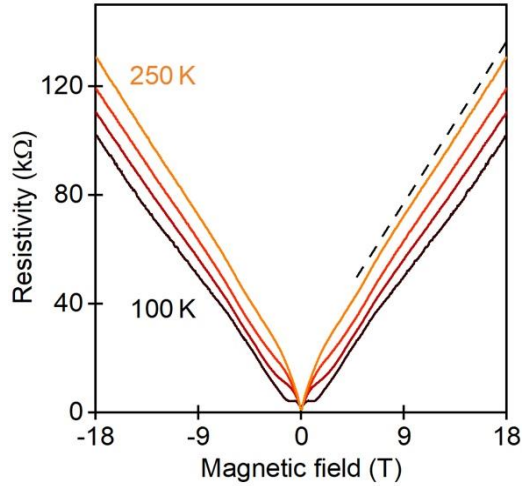

**Extended Data Fig. 5| Another example of quantum linear magnetoresistance.** Resistivity of MLG at the NP over a large range of $B$. Temperatures are between 100 and 250 K in steps of 50 K; device D2. Dashed line: Guide to the eye with a slope of 6.5 kOhm T$^{-1}$.

## 10. Landau quantization at room temperature

We have attributed the observed linear MR in high $B$ to the transition of the Dirac plasma into the quantized regime where the linear spectrum of MLG splits into dispersionless Landau levels. This condition is an essential prerequisite for discussing magnetotransport for the compensated Boltzmann gas on the zeroth LL. In MLG, the main cyclotron gap at the filling factor $\nu = 2$ is given by[87] $E[K] = v_F(2e\hbar B)^{1/2} \approx 400\times\sqrt{B[T]}$. The gap's size notably exceeds the thermal energy at room $T$ already in fields of a few T. Previously, the Landau quantization has been reported for ultrahigh magnetic fields of 30–40 T where even the quantum Hall effect was observed at room temperature[87]. To demonstrate that Landau quantization in our devices becomes important at room $T$ already in moderate $B$, Extended Data Fig. 6a shows the fan diagram measured for one of our Corbino devices at room $T$. The found peaks in inverse conductivity follow the main gaps at $\nu = \pm 2$, as expected, and become clearly visible at $B$ above 6 T. The Landau quantization is also visible in $\rho_{xx}$ measured in the standard Hall bar geometry (Extended Data Fig. 6b). These observations support the description of high-$B$ transport in neutral MLG in terms of the zeroth LL for the discussed temperature range up to 300 K.

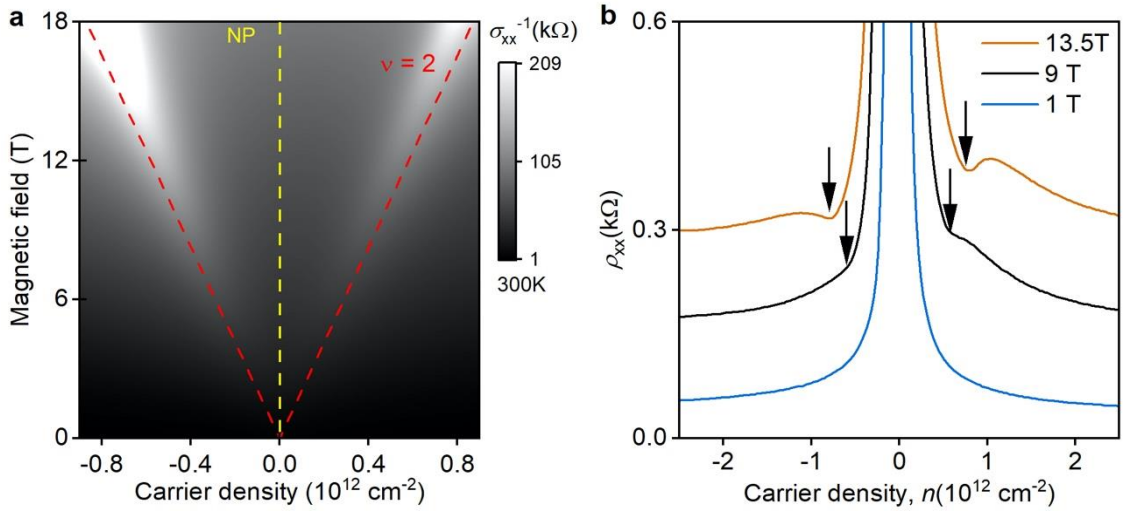

**Extended Data Fig. 6| Room-temperature Landau quantization in moderate magnetic fields. a**, Conductivity $\sigma_{xx}$ of MLG as a function of $B$ and carrier density at 300 K. The measurements are for a Corbino disk device. The vertical yellow line indicates the NP, and the red lines follow $\nu = \pm 2$. **b**, Room-$T$ resistivity for MLG measured in the Hall bar geometry at three representative $B$. The traces are shifted for clarity by 0.1 kΩ. The vertical arrows mark $\nu = \pm 2$.

## 11. Linear magnetoresistance in Corbino devices

We have also used our Corbino devices to rule out edge effects in the appearance of strange linear MR. Extended Data Fig. 7 shows that the linear dependence $\rho_{NP}(B)$ was also observed in this geometry, exhibiting little difference with respect to the behavior reported for the 4-probe Hall bar devices. Indeed, MR of Corbino disks is found to be weakly dependent on $T$ and exhibit slopes with values close to those observed in the Hall bar geometry (cf. Extended Data Fig. 5 and Fig. 3a of the main text). This proves that the linear magnetoresistance is an intrinsic (bulk) effect and, for example, it is not related to e-h annihilation at graphene edges[66] or to spin/valley Hall currents reported for neutral graphene[88].

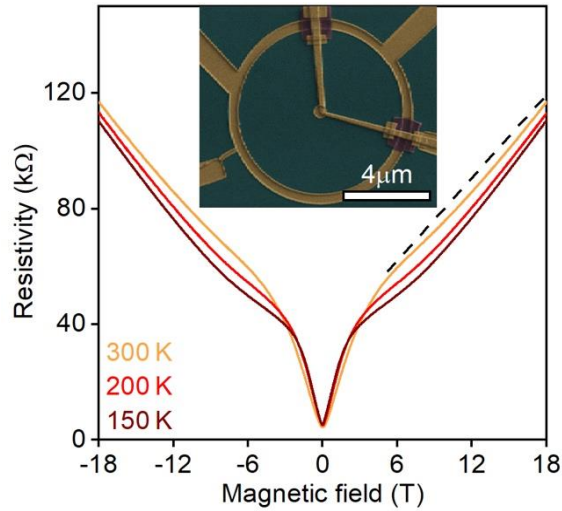

**Extended Data Fig. 7| Quantum linear magnetoresistance in Corbino devices.** $\rho_{NP}(B)$ measured for one of our Corbino-disk devices at different $T$ (color coded). The contact resistance was about 0.1 kOhm (measured in the limit of high $n$ and assumed to change little near the NP). Black line: guide to the eye with the slope 4.8 kΩ T$^{-1}$. Insert: false color micrograph of the device. Green, hBN on top of graphene; gold, metallic contacts; purple, polymer bridges over the outer ring contact, which are required for the metallization to reach the inner contact.

## 12. Magnetoresistance of low-mobility graphene

To illustrate the importance of high quality for the reported magnetoresistance behavior of MLG in both low and high $B$, we have measured low-mobility devices obtained by exfoliation of graphene onto an oxidized Si wafer (inset of Extended Data Fig. 8a). At liquid-helium $T$, such devices exhibited strong charge inhomogeneity with $\delta n \approx 10^{11}$ cm$^{-2}$ (Extended Data Fig. 8a) which was nearly two orders of magnitude higher than that for hBN-encapsulated graphene (Fig. 1c of the main text). Even at 300 K, thermally excited density $n_{th}$ remained smaller than the residual $\delta n$, which means that electron transport near the NP in such devices was dominated by charge inhomogeneity (e-h puddles) at all $T$ in the experiment. Accordingly, although $\rho_{NP}$ decreased with increasing $T$ (Extended Data Fig. 8), similar to the case of our high-mobility devices, it only reached ~4 kOhm at room $T$, significantly away from the intrinsic value of ~1 kOhm for the Dirac plasma in the quantum-critical regime.

In small magnetic fields, $\rho_{NP}$ for MLG-on-SiO$_2$ evolved quadratically with $B$ (Extended Data Fig. 8b). The measured $\Delta$ was found more than two orders of magnitude smaller than in high-quality MLG (< 1% at 0.1 T), which corresponds to ~8,500 cm$^2$ V$^{-1}$ s$^{-1}$ at the NP. With increasing $B$ above 1 T, MR of graphene on SiO$_2$ deviated from the parabolic dependence and became sublinear at high $T$ (Extended Data Fig. 8c), in agreement with the previous reports[30,70]. Such sublinear behavior may be attributed to short-range scattering[73], which is present in graphene-on-SiO$_2$[89], but further research is required to unambiguously identify origins of high-$B$ MR in low-mobility MLG. Nonetheless, our observations clearly show the importance of electronic quality for the observation of the linear magnetoresistivity.

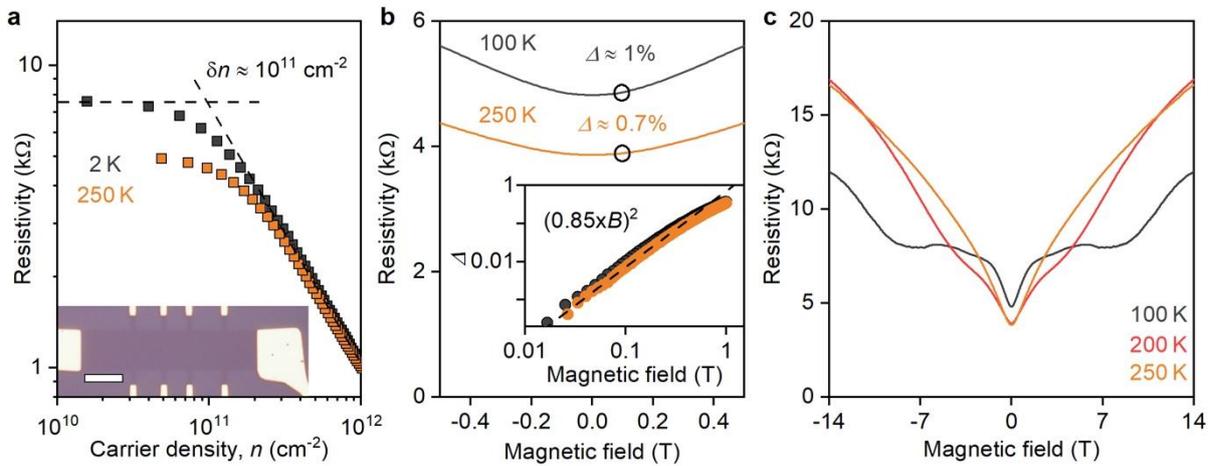

**Extended Data Fig. 8| Magnetotransport in graphene-on-silicon-oxide. a**, Resistivity at the NP as a function of $n$ plotted for two characteristic $T$. The crossing of the two dashed lines indicates the charge inhomogeneity level. The inset shows an optical image of the studied device. Scale bar, 10 μm. **b**, Resistivity at the NP as a function of $B$. The open circles mark MR values at 0.1 T. Inset: same curves replotted on a log scale. The dashed line is the parabolic fit for the 250-K curve below 0.2 T. **c**, Resistivity of charge-neutral graphene-on-SiO$_2$ over a large range of $B$ at different $T$. No clear linear MR is observed at any $T$ for such MLG devices.

## Method References.

**Acknowledgements.** We acknowledge financial support from the European Research Council (grant VANDER), the Lloyd's Register Foundation, Graphene Flagship Core3 Project; J.L. and A.P. were supported by the EU Horizon 2020 programme under the Marie Skłodowska-Curie grants 891778 and 873028, respectively; V.I.F also acknowledges the support from EPSRC grants EP/W006502/1, EP/V007033/1 and EP/S030719/1; A.P. and A.K. acknowledge support from the Leverhulme Trust (grant RPG-2019-363).

**Author contributions.** A.I.B., L.A.P. and A.K.G. designed and supervised the project; N.X., P.K., Z.W. fabricated the graphene devices; A. M. provided multilayer graphene devices; J.L., A.I.B., L.A.P., J.B., C.M. carried out the electrical measurements; A.I.B., J.L., L.A.P. and A.K.G. analyzed data with help from V.I.F, A.P., A.E.K., A.A.G., N.X., P.K.; A.K.G and A.I.B. wrote the manuscript with contributions from I.V.G., A.P., A.E.K. and V.I.F. All authors contributed to discussions.